\documentclass[
reprint,
superscriptaddress,
amsmath,amssymb,
aps,
]{revtex4-2}

\usepackage[version=4]{mhchem}
\usepackage{siunitx}
\usepackage{miller}
\usepackage{graphicx}
\usepackage{dcolumn}
\usepackage{bm}
\usepackage[newcommands]{ragged2e}
\usepackage{caption}
\usepackage[colorlinks, linkcolor=blue, anchorcolor=blue, citecolor=blue]{hyperref}

\begin{document}

\title{Deep Learning Illuminates Spin and Lattice Interaction in Magnetic Materials}

\author{Teng Yang}\thanks{these authors contribute equally}
\affiliation{Graduate School of China Academy of Engineering Physics, Beijing 100088, P. R. China}
\affiliation{School of Materials Science and Engineering, Tsinghua University, Beijing 100084, P. R. China}

\author{Zefeng Cai}\thanks{these authors contribute equally}
\affiliation{School of Materials Science and Engineering, Tsinghua University, Beijing 100084, P. R. China}

\author{Zhengtao Huang}\thanks{these authors contribute equally}
\affiliation{Graduate School of China Academy of Engineering Physics, Beijing 100088, P. R. China}
\affiliation{International School of Materials Science and Engineering, \\ Wuhan University of Technology, Wuhan 430070, P. R. China}

\author{Wenlong Tang}
\affiliation{Graduate School of China Academy of Engineering Physics, Beijing 100088, P. R. China}

\author{Ruosong Shi}
\affiliation{Graduate School of China Academy of Engineering Physics, Beijing 100088, P. R. China}

\author{Andy Godfrey}
\affiliation{School of Materials Science and Engineering, Tsinghua University, Beijing 100084, P. R. China}

\author{Hanxing Liu}
\affiliation{International School of Materials Science and Engineering, \\ Wuhan University of Technology, Wuhan 430070, P. R. China}

\author{Yuanhua Lin}
\affiliation{School of Materials Science and Engineering, Tsinghua University, Beijing 100084, P. R. China}

\author{Ce-Wen Nan}
\affiliation{School of Materials Science and Engineering, Tsinghua University, Beijing 100084, P. R. China}

\author{Meng Ye}
\affiliation{Graduate School of China Academy of Engineering Physics, Beijing 100088, P. R. China}

\author{LinFeng Zhang}
\affiliation{AI for Science Institute, Beijing 100080, P. R. China}

\author{Han Wang}
\email{wang_han@iapcm.ac.cn}
\affiliation{Laboratory of Computational Physics, Institute of Applied Physics and Computational Mathematics, \\ Fenghao East Road 2, Beijing 100094, P.R. China}
\affiliation{HEDPS, CAPT, College of Engineering, Peking University, Beijing 100871, P.R. China}

\author{Ben Xu}
\email{bxu@gscaep.ac.cn}
\affiliation{Graduate School of China Academy of Engineering Physics, Beijing 100088, P. R. China}

\date{\today}

\begin{abstract}
Atomistic simulations hold significant value in clarifying crucial phenomena such as phase transitions and energy transport in materials science. Their success stems from the presence of potential energy functions capable of accurately depicting the relationship between system energy and lattice changes. In magnetic materials, two atomic scale degrees of freedom come into play: the lattice and the spin. However, accurately tracing the simultaneous evolution of both lattice and spin in magnetic materials at an atomic scale is a substantial challenge. This is largely due to the complexity involved in depicting the interaction energy precisely, and its influence on lattice and spin-driving forces, such as atomic force and magnetic torque, which continues to be a daunting task in computational science. Addressing this deficit, we present DeepSPIN, a versatile approach that generates high-precision predictive models of energy, atomic forces, and magnetic torque in magnetic systems. This is achieved by integrating first-principles calculations of magnetic excited states with deep learning techniques via active learning. We thoroughly explore the methodology, accuracy, and scalability of our proposed model in this paper. Our technique adeptly connects first-principles computations and atomic-scale simulations of magnetic materials. This synergy presents opportunities to utilize these calculations in devising and tackling theoretical and practical obstacles concerning magnetic materials.
\end{abstract}

\maketitle

Spin-lattice interaction describes the coupling between the magnetic and lattice subsystems in magnetic materials, which is pivotal for understanding phase transitions and transport behaviors. Additionally, it profoundly impacts various advanced research areas in magnetic materials, including antiferromagnetic spintronics \cite{RevModPhys.90.015005}, terahertz spintronics \cite{Nova_NP_2017}, spin-caloritronics \cite{Seki_PRL_2015}, curvilinear magnetism \cite{Kravchuk_PRL2018}, and even quantum criticality \cite{Narayan_NM_2019}. Given its significance, there is a pressing need for accurate descriptions of the coupling between non-collinear magnetic configurations and intricate atomic structures at an atomic resolution. Moreover, to study transport properties effectively, it is essential to precisely predict the driving forces that influence the evolution of atomic positions and magnetic moments. However, conventional computational approaches have their limitations. For instance, first-principles methods are constrained by scale, while phase field methods can lack in spatial and energy resolution. Current atomic-scale simulation techniques, such as molecular dynamics, micromagnetics, and Monte Carlo, typically focus on the evolution of only one degree of freedom, either the lattice or the spin. 

In recent years, significant progress has been made in the development method cope both spin and lattice degree of freedom(d.o.f.), aiming to couple molecular dynamics with spin dynamics at the atomic scale \cite{Bellaiche_2020, MPW_PRB2008, MPW_CPC2016, PD_PRB2017, tranchida2018massively}. While these methods include equations for evolving both atomic and magnetic moments, applying these methods still necessitates potential functions that accurately capture both spin and lattice interactions. Researchers have approached this challenge from both analytical and numerical perspectives. Analytically, many opt for the Bethe-Slater function \cite{tranchida2018massively, Nikolov2021}. However, its application necessitates individual parameter fittings for each material, based on computational or experimental values. Additionally, its spherically symmetric form limits consideration to the influence of radial atom distances on spin-lattice coupling. Numerically, efforts have been made to offset this limitation by calculating the first and second energy derivatives concerning spins and atomic coordinates \cite{MFechner_PRB_2018}. These techniques tend to work best when the effect of spin variation on energy is subtle, struggling with intricate lattice and spin configurations or phase transitions. Although machine learning has recently made inroads in this domain, it's not without its challenges. Some methods segregate lattice \cite{ZLF_PRL2018} and spin \cite{li2022deeplearning} contributions to energy, using handcrafted magnetic Hamiltonians \cite{tranchida2018massively}, which can introduce double counting issues. Others focus merely on collinear \cite{novikov2022magnetic} or spiral \cite{tranchida2018massively} magnetic configurations and the lattice's volumetric effect on the spin system \cite{Nikolov2021}, neglecting arbitrary non-collinear excited states. Furthermore, current machine learning approaches fall short in predicting the torque, or the magnetic force, propelling spin evolution.

In this work, we develop a general model based on deep learning to predict the energy and driving forces of non-collinear magnetic configurations and complex lattice structures, aiming to reveal the intricate spin-lattice coupling in magnetic materials. Initially, we employ a novel ``pseudo-atom" approach to map the connections between lattice and spin configurations as interdependent features. Subsequently, we design the specific structure of deep neural networks and the loss function for model training to capture these interactions effectively. By implementing an active learning strategy, we obtain a cost-minimized dataset that enhances the model's performance. The potential energy surface and its second-order differential properties (such as phonon and magnon spectra) demonstrate the remarkable accuracy of the model in describing spin-lattice interactions at various energy scales. The model exhibits strong generalizability for large systems and those with drastic changes in lattice and spin configurations, as evidenced by its predictions for the potential barrier in the multiferroic BiFeO$_3$ polarization reversal and the magnetic structures on high-angle symmetric grain boundaries in antiferromagnetic NiO.

\begin{figure*}[t!]
    \centering
    \includegraphics[width=0.8\linewidth]{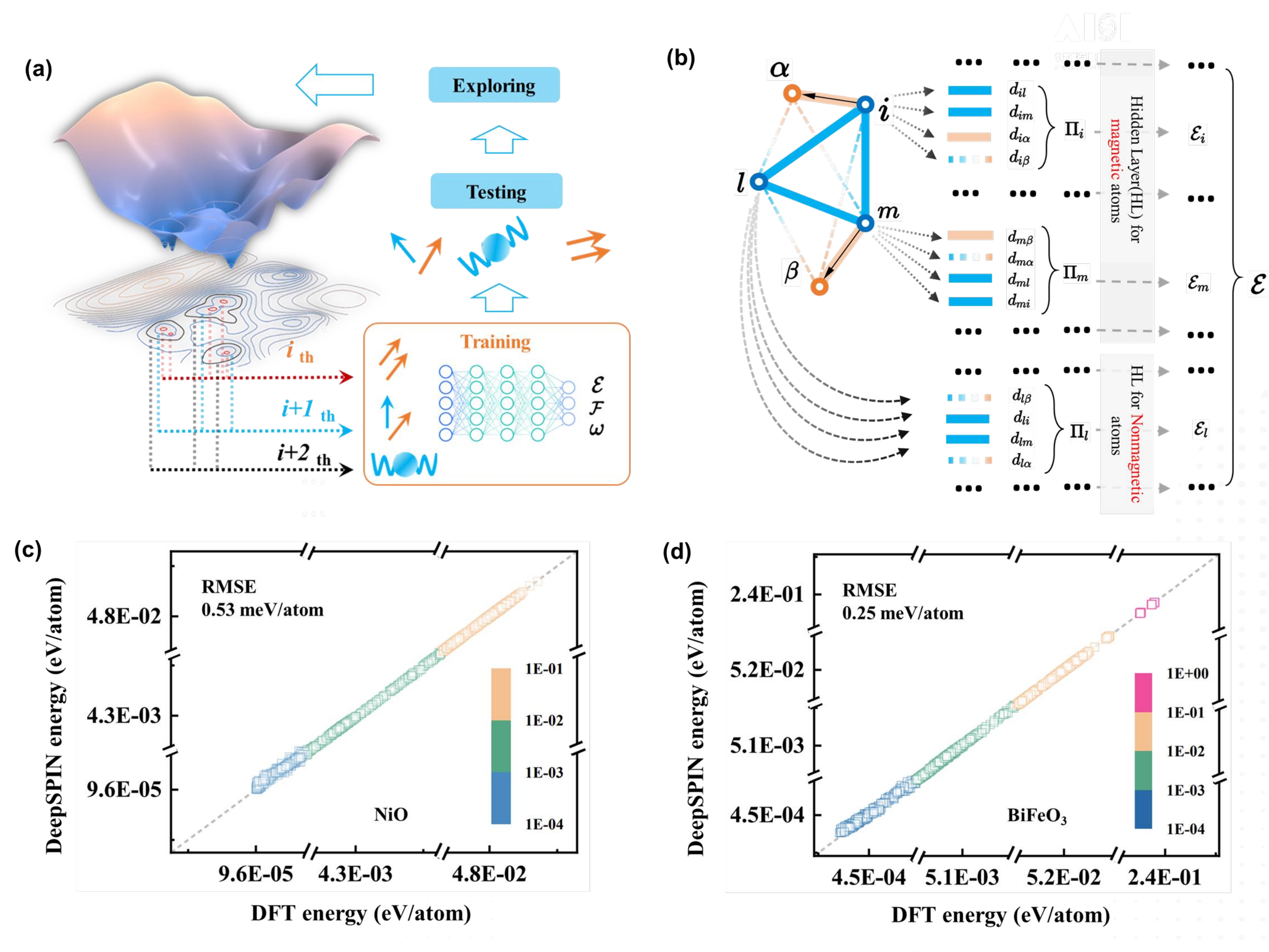}
    \caption{Active learning diagram to obtain DeepSPIN models. (a) Iterative sampling of the potential energy surface in order of the configuration's energy scale. Each iteration consists of training DeepSPIN models from the current dataset, testing the models' accuracy, and exploring the configuration space for the next energy scale utilizing active learning. (b) Schematic diagram of constructing the DeepSPIN model with pseudo-atoms. $i$, $m$ and $l$ are real atoms, $\alpha$ and $\beta$ are pseudo-atoms for example. The local environment matrix $\bm{\Pi}_i$ consists of the coordinate information from all neighboring atoms, and is mapped to atomic energy $\mathcal{E}_i$ by neural networks. (c)(d) The comparisons of energies predicted by DeepSPIN and labelled from DFT on NiO and BiFeO$_3$, respectively. RMSE of predictions is also shown.}
    \label{fig:active} 
\end{figure*}

We develop a high-precision model capable of accurately describing the potential energy surface $\mathcal{E}\left(\mathcal{R}, \mathcal{S}\right)$ corresponding to arbitrary lattice configurations $\mathcal{R}$ and spin configurations $\mathcal{S}$. For each magnetic real atom $i$, we introduce an associated massless ``pseudo-atom" $i^p$ (e.g., atoms $\alpha$ and $\beta$ in Fig.~\ref{fig:active}(b)) around it, representing the spin $\bm{S}_i$ of atom $i$ through the Cartesian coordinates $\bm{R}_{i^p}$ of pseudo-atom $i^p$, i.e., $\bm{R}_{i^p} = \bm{R}_{i} + \eta_{\zeta_i} \cdot \bm{S}_i$, where the hyperparameter $\eta_{\zeta_i}$ determines the distance between atom $i$ and $i^p$ based on the element type $\zeta_i$ of atom $i$. We enumerate all atom indices $j$ satisfying $\Vert\bm{r}_{ij}\Vert < r_c$ (regardless of whether atom $j$ is a real atom or a pseudo-atom) to generate a neighbor list $\Omega\left(i\right)$ for each real atom $i$, where $\bm{r}_{ij} = \bm{R}_j - \bm{R}_i$ represents the relative coordinate vector, and $r_c$ is the cutoff radius. Following the DeepPot-SE scheme \cite{zhang2018deep, zhang2018end, wang2018deepmd}, we denote the cardinality of $\Omega\left(i\right)$ as $N_i$ and construct the local environment matrix $\bm{\Pi}_i \in \mathbb{R}^{N_i \times 4}$ for each real atom $i$, where each row $\bm{d}_{ij}$ contains the local coordinate information of the neighboring atom $j$,
\begin{equation}
\bm{d}_{ij} = \left\{\frac{s\left(r_{ij}\right)}{r_{ij}}, \frac{s\left(r_{ij}\right)x_{ij}}{r_{ij}^2}, \frac{s\left(r_{ij}\right)y_{ij}}{r_{ij}^2}, \frac{s\left(r_{ij}\right)z_{ij}}{r_{ij}^2}\right\}.
\end{equation}

Here $r_{ij} = \Vert \bm{r}_{ij} \Vert$, $x_{ij}$, $y_{ij}$, and $z_{ij}$ are the three components of $\bm{r}_{ij}$. The smooth factor $s\left(r_{ij}\right)$ is used to ensure the numerical continuity at the cutoff boundary. It is noteworthy that $\bm{\Pi}_i$ naturally encompasses three different types of interactions in magnetic systems: lattice-lattice interaction, manifested as the positional relationship between atom $i$ and its neighboring real atoms (e.g., $\bm{d}_{im}$ in Fig.~\ref{fig:active}(b)); lattice-spin interaction, represented by the position relationship between atom $i$ and its neighboring pseudo-atoms (e.g., $\bm{d}_{i\beta}$ in Fig.~\ref{fig:active}(b)); and spin-spin interaction, expressed by the position relationship between the pseudo-atom $i^p$ and its neighboring pseudo-atoms (e.g., $\bm{d}_{i\alpha}$ and $\bm{d}_{i\beta}$ in Fig.~\ref{fig:active}(b)).

Next, we adopt a deep neural network \cite{goodfellow2016deep} to take each $\bm{\Pi}_i$ as input and output the corresponding local atomic energy $\mathcal{E}_i$. The neural network comprises two parts \cite{zhang2018end}: the embedding network, which is a specially designed network that encodes $\bm{\Pi}_i$ into high-dimensional feature vectors preserving the system's translational, rotational, and permutation symmetries; and the fitting network, which is a fully connected residual network \cite{he2016deep} that maps the obtained feature vectors to $\mathcal{E}_i$. The network parameters corresponding to each element type are independent of each other and shared among all atoms belonging to that type. The total energy $\mathcal{E}$ of the system is expressed as the sum of atomic energies $\mathcal{E}_i$ for all $N$ real atoms, i.e., $\mathcal{E} = \sum_{i}^N {\mathcal{E}_i}$, thereby preserving the extensive character.

The atomic force $\mathcal{F}_i$ can be analytically expressed as the derivative of $\mathcal{E}$ with respect to the atomic position $\bm{R}_i$ and expanded by the chain rule, as shown in Eq.\eqref{eq:atomic_force}, where $\Omega_r\left(i\right)$ represents all neighboring real atoms of atom $i$. Since the position $\bm{R}_i$ and the corresponding pseudo-atom position $\bm{R}_{i^p}$ jointly affect $\mathcal{E}_j$, we introduce the coefficient $\delta_i$ to distinguish the magnetism of atoms, i.e., $\delta_i = 1$ for magnetic atoms and $\delta_i = 0$ otherwise. Similarly, the magnetic torque $\omega_i$ can be expressed as the derivative of $\mathcal{E}$ with respect to the spin $\bm{S}_i$, as shown in Eq.\eqref{eq:magnetic_torque}. Here $\bm{S}_i$ affects $\mathcal{E}_j$ through the pseudo-atom position $\bm{R}_{i^p}$. In this way, both atomic forces and magnetic torques are influenced by real and pseudo atoms.
\begin{align}
\mathcal{F}_i &= -\frac{\partial \mathcal{E}}{\partial \bm{R}_i} = -\sum_k^N \frac{\partial \mathcal{E}_k}{\partial \bm{\Pi}_k} \cdot \frac{\partial \bm{\Pi}_k}{\partial \bm{R}_i} \nonumber \\
&= -\sum_{j \in \Omega\left(i\right)} \frac{\partial \mathcal{E}_i}{\partial \bm{d}_{ij}} \cdot \frac{\partial \bm{d}_{ij}}{\partial \bm{R}_i} \nonumber \\
&+ \sum_{j \in \Omega_r\left(i\right)} \left\{ \frac{\partial \mathcal{E}_j}{\partial \bm{d}_{ji}} \cdot \frac{\partial \bm{d}_{ji}}{\partial \bm{R}_j} + \delta_i \cdot \frac{\partial \mathcal{E}_j}{\partial \bm{d}_{ji^p}} \cdot \frac{\partial \bm{d}_{ji^p}}{\partial \bm{R}_j} \right\},
\label{eq:atomic_force} 
\end{align}
\begin{align}
\omega_i &= -\frac{\partial \mathcal{E}}{\partial \bm{S}_i} = -\sum_k^N \frac{\partial \mathcal{E}_k}{\partial \bm{\Pi}_k} \cdot \frac{\partial \bm{\Pi}_k}{\partial \bm{R}_{i^p}} \cdot \frac{\partial \bm{R}_{i^p}}{\partial \bm{S}_i} \nonumber \\
&= \left\{ \frac{\partial \mathcal{E}_i}{\partial \bm{d}_{ii^p}} \cdot \frac{\partial \bm{d}_{ii^p}}{\partial \bm{R}_i} + \sum_{j \in \Omega_r\left(i\right)} \frac{\partial \mathcal{E}_j}{\partial \bm{d}_{ji^p}} \cdot \frac{\partial \bm{d}_{ji^p}}{\partial \bm{R}_j} \right\} \cdot \eta_{\zeta_i}.
\label{eq:magnetic_torque} 
\end{align}

\begin{figure}[t!]
    \centering
    \includegraphics[width=1.0\linewidth]{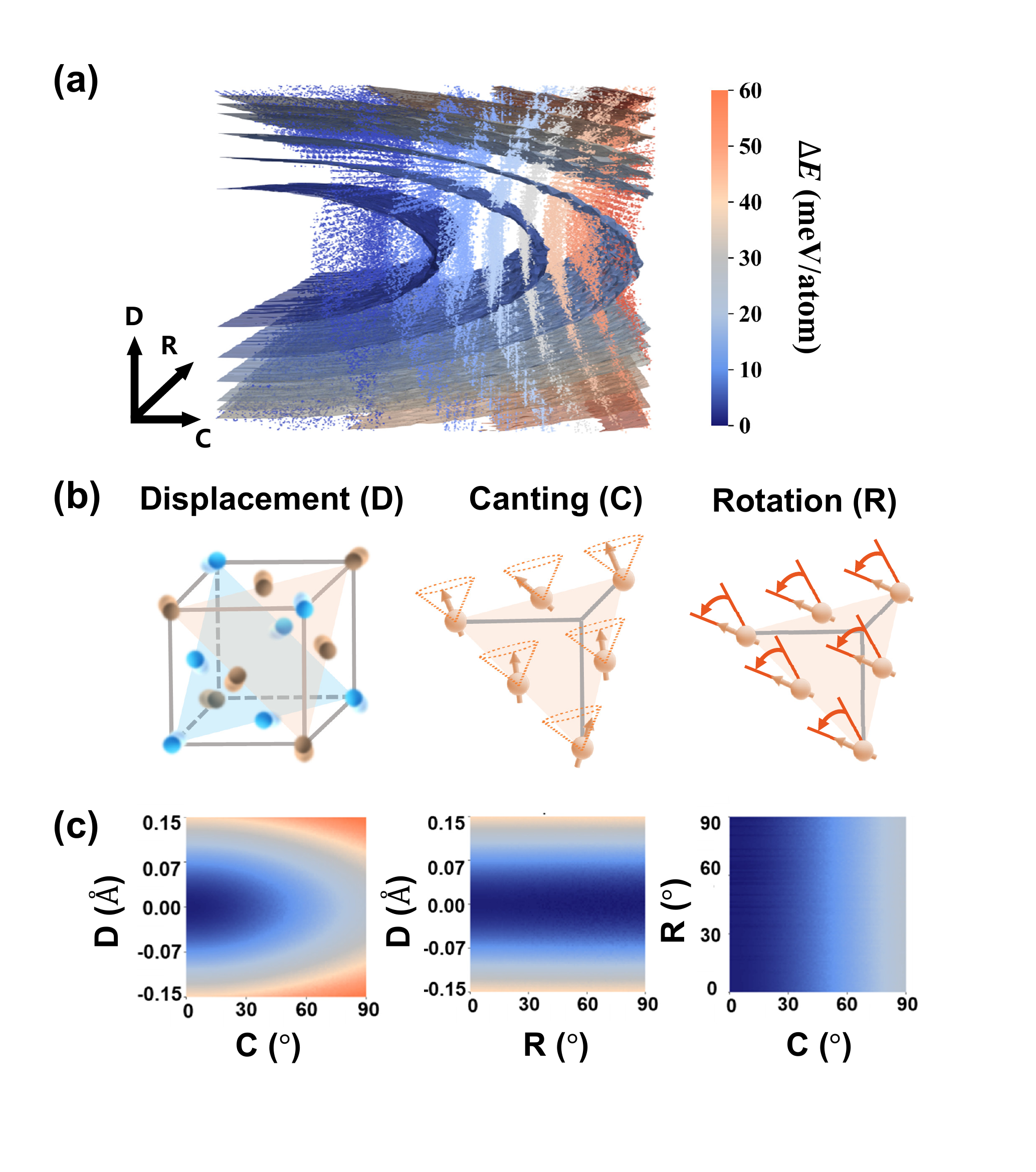}
    \caption{Profiles of high dimensional potential energy surface (PES). (a) 3-dimensional PES of NiO constructed with respect to diverse physical events, varying colors represent different energy scale relative to the ground state. (b) Illustrations of 3 distinct physical events, including atomic displacement, spin rotation, and spin canting. (c) Projections of PES along each axis, offering the quantitative assessment of energy from different pertubations.}
    \label{fig:NiO_PES}
\end{figure}

To efficiently train the neural network, we design the loss function in the form of Eq.\eqref{eq:loss_func}, where $\Delta$ represents the difference between the DeepSPIN prediction and the label, $N$ and $N_s$ represent the number of real atoms and pseudo-atoms respectively, and $p_\mathcal{E}$, $p_\mathcal{F}$, and $p_{\omega}$ are adjustable weights controlling the contributions of atomic energy, atomic forces, and magnetic torques in the loss function respectively. We employ the Adam optimizer \cite{kingma2017adam} to minimize the loss function, ensuring accurate predictions for each component and achieving faster training speed.
\begin{align}
&\mathcal{L}\left(p_\mathcal{E}, p_\mathcal{F}, p_{\omega}\right) = \nonumber \\
&\quad p_\mathcal{E}\left( \frac{\Delta\mathcal{E}}{N}\right)^2 + \frac{p_\mathcal{F}}{3N}\sum_i^N{\Vert\Delta\mathcal{F}_i\Vert^2} + \frac{p_{\omega}}{3N_s}\sum_i^{N_s}{\Vert\Delta\mathcal{\omega}_i\Vert^2}.
\label{eq:loss_func}
\end{align}
Moreover, to obtain high-precision labels, we perform first-principles calculations on non-collinear magnetic configurations using the DeltaSpin scheme \cite{cai2023firstprinciples}. DeltaSpin optimizes the Lagrangian function $\mathcal{L}\left[\rho; \left\{\bm{\lambda}_i, \bm{S}_i, \bm{S}_i^{*}\right\}\right] = E_{\mathrm{KS}}\left[\rho\right] - \sum_i\bm{\lambda}_i \cdot \left(\bm{S}_i\left[\rho\right] - \bm{S}_i^{*}\right)$, which can lead to the errors of magnetic moments and energy converging to $\delta\bm{S} = \num{e-5}\ \mu_B$ and $\delta\mathcal{E} = \num{e-9}\ \mathrm{eV}$ respectively. In this approach, magnetic torque $\omega_i$ can be obtained using a method similar to the Hellman-Feynman scheme \cite{feynman1939forces}, i.e., $\omega_i = -\delta\mathcal{L} / \delta\bm{S}_i^* = -\bm{\lambda}_i$.

Starting from the magnetic ground state, we employ an active learning strategy \cite{zhang2019active, zhang2020dpgen} to explore the configuration space of lattice and spin, and construct the system's potential energy surface (PES). Following the order of increasing energy scales for excited states, we design three distinct physical events for the ground state and perform random sampling. These events involve perturbing the positions of real atoms (Displacement, abbreviated as D), changing the angles between spins (Canting, abbreviated as C), or simultaneously altering the orientations of all spins (Rotation, abbreviated as R), as illustrated in Fig.~\ref{fig:NiO_PES}(b). The dataset expansion is accomplished iteratively through alternating sampling and model training. In each training round, we train four models with different random seeds and use the prediction deviations of atomic forces and magnetic torques on newly generated configurations as filtering criteria. Only new configurations meeting the criteria can be added to the dataset, thereby excluding unphysical configurations with excessively high energy or configurations sufficiently learned \cite{zhang2020dpgen}. Through multiple iterations, the upper limits for the physical events can be gradually increased to 0.15 \AA, $180^{\circ}$, and $180^{\circ}$, respectively, while ensuring minimization of the dataset expansion cost.

\begin{figure}[t!]
    \centering
    \includegraphics[width=1.0\linewidth]{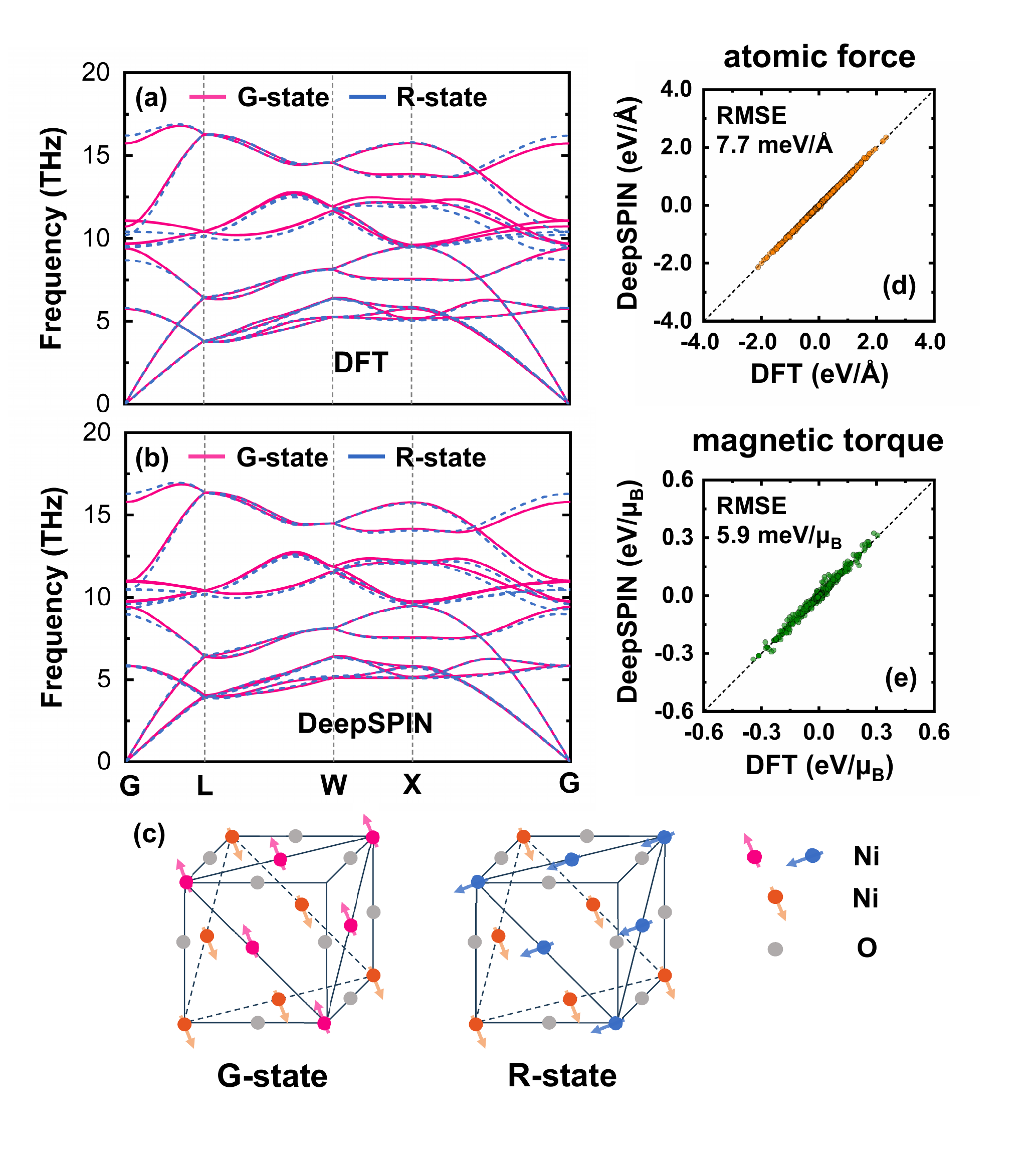}
    \caption{Validating predictions of spin-lattice coupling via lattice dynamics analysis. (a)(b) The comparisons of NiO phonon spectra of G-state (red) and R-state (blue) configuration, obtained from DFT and DeepSPIN respectively. (c) Illustrations of G-state and R-state, representing in-plane spin rotation of Ni atoms. (d)(e) Comparisons of atomic forces and magnetic torques predicted by DeepSPIN and labelled from DFT, respectively. RMSE is shown.}
    \label{fig:Phonon}
\end{figure}

We highlight the efficacy of DeepSPIN method through studies on two antiferromagnetic insulating materials: NiO and BiFeO$_3$. NiO exhibits stable antiferromagnetic order with minimal magnetic anisotropy \cite{milano2010magnetic, massey1990pressure, aytan2017spin}. In contrast, BiFeO$_3$ is a multiferroic system characterized by pronounced Dzyaloshinskii-Moriya (DM) interactions, embodying a complex interplay between oxygen octahedral rotations and subdued ferromagnetic moments \cite{fischer1980temperature, ederer2005weak, albrecht2010ferromagnetism, xu2021first}. Fig.~\ref{fig:active}(c) and (d) depict a compelling agreement between the energies predicted by the DeepSPIN model and those derived from DFT for excited states. Notably, the distribution of average relative atomic energy spans four orders of magnitude from \num{e-4}~eV to \num{e-1}~eV, represented by varying colors, underscoring the model's impressive accuracy. In Fig.~\ref{fig:NiO_PES}(c), we present the delineated potential energy surface profiles for three unique physical events. This visualization allows for nuanced evaluations of energy associated with various excitation modes. Fig.~\ref{fig:Phonon}(c) and (d) further showcase the model's proficiency in predicting atomic forces and magnetic torques, registering RMSE values of $5.9$~meV/$\mu$B and $7.7$~meV/\AA, respectively. The precise torque predictions can be seamlessly integrated into frameworks like time-dependent DFT (TD-DFT) \cite{vignale1996current, qian2002dynamical} or Landau-Lifshitz-Gilbert (LLG) equations \cite{gilbert2004phenomenological}, serving as primary drivers for spin evolution.

The DeepSPIN model can also reveal the influence of spin-lattice interaction on the dynamic properties of magnetic systems, such as phonon spectra and magnon spectra \cite{kumar2012spin, wu2018magnon, hellsvik2019general}. Fig.~\ref{fig:Phonon}(a) and (b) illustrate the impact of varying magnetic configuration on lattice dynamics. When the magnetic configuration of NiO changes from the antiferromagnetic ground state (G-state) to mutually orthogonal excited states (R-state), as shown in Fig.~\ref{fig:Phonon}(c), acoustic branches of the phonon spectrum remain nearly unchanged, while the frequencies of the optical branches undergo significant variations with magnitude close to 3 meV. This reflects the distinct interactions between magnetic Ni atoms within or between the $\lbrace111\rbrace$ plane as well as the super-exchange interaction along the Ni-O-Ni chain \cite{massey1990pressure}. What's more, the impact of lattice configurations on spin dynamics can be also revealed (see Fig. S4 in the SM). When $-3\%$ uniform compressive strain is applied to $R3c$ BiFeO$_3$, the magnon frequency increases significantly, indicating enhanced magnetic interactions and greater stability of the antiferromagnetic order. The comparison of the spectra obtained from DeepSPIN with DFT calculations highlights the accuracy of the DeepSPIN model.

\begin{figure*}[t!]
    \centering
    \includegraphics[width=0.8\linewidth]{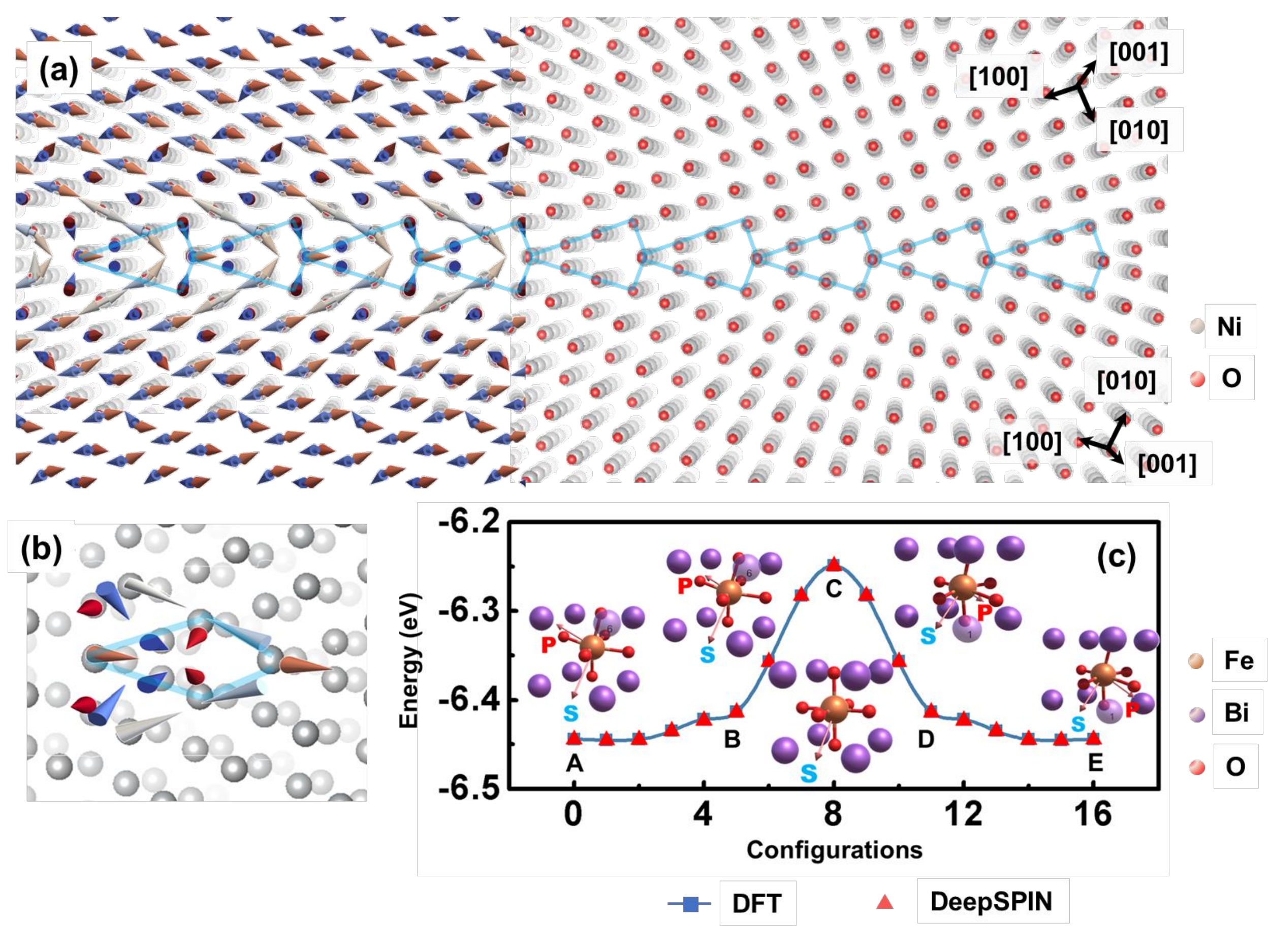}
    \caption{Predictions of spin-lattice coupling in complex crystal structures. (a) The optimized spin configuration within $\Sigma$5 symmetric NiO grain boundary (left side) and the crystal structure obtained from high-resolution electron microscopy \cite{merkle1987atomic} (right side). (b) Enlarged view of the local Ni spin distribution on the grain boundaries. Varying color indicates y-axis component magnitude, with blue for [112] and red for [$\overline{1}$$\overline{1}$$\overline{2}$]. (c) Energy barrier derived from DFT and DeepSPIN for the \ang{180} polarization switching trajectory in BiFeO$_3$ with varying polarization $\bm{P}$ and fixed magnetization $\bm{S}$. The modulus of P gradually diminishes to 0 and then increases inversely.}
    \label{fig:GB}
\end{figure*}

DeepSPIN demonstrates superior generalizability in various application scenarios. We construct a high-angle NiO grain boundary system ($\Sigma$5) with mirror-symmetric atomic distribution on both sides of the grain boundary $\left(\overline{3}\overline{1}\overline{0}\right)$ \cite{merkle1987atomic}. All Ni atoms are initially set to antiferromagnetic ground state. The system contains 11,328 atoms in total and has dimensions of 8 nm, 1.6 nm, and 7.7 nm along three directions, exceeding the affordable scale of first-principles calculations. Fig.~\ref{fig:GB}(a) and (b) illustrate the optimization results of the spin configuration by the DeepSPIN model. The Ni atoms maintain the $\langle11\overline{2}\rangle$ antiferromagnetic order away from the grain boundary, but exhibit highly irregular distorted configurations near the boundary, resulting in local net magnetic moments, consistent with previous studies \cite{li2006magnetic}. Additionally, we apply the nudge elastic band (NEB) method \cite{henkelman2000climbing, munro2018discovering} to calculate the \ang{180} polarization switching process of $R3c$ BiFeO$_3$ along the $\left\lbrack111\right\rbrack$ direction \cite{zavaliche2005polarization}. As shown in Fig.~\ref{fig:GB}(c), DeepSPIN model not only successfully obtains the energy barrier for polarization switching but also accurately predicts the energy of the distorted structures along the switching trajectory \cite{heron2014deterministic}. This indicates DeepSPIN is effectively capable of capturing the complex spin-lattice interaction involving Bi displacements and FeO$_6$ oxygen octahedral distortions.

In conclusion, we propose a flexible approach to derive spin-lattice coupling models, which facilitates the construction of deep neural networks for magnetic materials' energy, atomic forces, and magnetic torques. This method outperforms other electronic and mesoscale techniques when dealing with various magnetic and crystal structures, especially those exhibiting significant irregularities in lattice and spin arrangements. DeepSPIN notably maintains high accuracy even when the average atomic energy reaches \SI{e-4}{eV}, atomic forces below 10 meV/\AA, and magnetic torques below 10 meV/$\mu_B$. Although the ``curse of dimensionality" requires a relatively large dataset in configuration space to achieve the aforementioned accuracy, this difficulty can be mitigated by the active learning scheme. Through integrating DeepSPIN with atomic-scale spin-lattice dynamics techniques, we are now equipped to address various phenomena, such as paramagnetic states, phonon-magnon interactions, and even quantum critical phenomena in magnetic materials.

B.X. acknowledges the support of the National Natural Science Foundation of China (NSFC, Grant No. 52072209), the Basic Science Center Program of the National Natural Science Foundation of China (Grant No. 52388201).

\bibliography{main}

\begin{thebibliography}{44}%
\makeatletter
\providecommand \@ifxundefined [1]{%
 \@ifx{#1\undefined}
}%
\providecommand \@ifnum [1]{%
 \ifnum #1\expandafter \@firstoftwo
 \else \expandafter \@secondoftwo
 \fi
}%
\providecommand \@ifx [1]{%
 \ifx #1\expandafter \@firstoftwo
 \else \expandafter \@secondoftwo
 \fi
}%
\providecommand \natexlab [1]{#1}%
\providecommand \enquote  [1]{``#1''}%
\providecommand \bibnamefont  [1]{#1}%
\providecommand \bibfnamefont [1]{#1}%
\providecommand \citenamefont [1]{#1}%
\providecommand \href@noop [0]{\@secondoftwo}%
\providecommand \href [0]{\begingroup \@sanitize@url \@href}%
\providecommand \@href[1]{\@@startlink{#1}\@@href}%
\providecommand \@@href[1]{\endgroup#1\@@endlink}%
\providecommand \@sanitize@url [0]{\catcode `\\12\catcode `\$12\catcode
  `\&12\catcode `\#12\catcode `\^12\catcode `\_12\catcode `\%12\relax}%
\providecommand \@@startlink[1]{}%
\providecommand \@@endlink[0]{}%
\providecommand \url  [0]{\begingroup\@sanitize@url \@url }%
\providecommand \@url [1]{\endgroup\@href {#1}{\urlprefix }}%
\providecommand \urlprefix  [0]{URL }%
\providecommand \Eprint [0]{\href }%
\providecommand \doibase [0]{https://doi.org/}%
\providecommand \selectlanguage [0]{\@gobble}%
\providecommand \bibinfo  [0]{\@secondoftwo}%
\providecommand \bibfield  [0]{\@secondoftwo}%
\providecommand \translation [1]{[#1]}%
\providecommand \BibitemOpen [0]{}%
\providecommand \bibitemStop [0]{}%
\providecommand \bibitemNoStop [0]{.\EOS\space}%
\providecommand \EOS [0]{\spacefactor3000\relax}%
\providecommand \BibitemShut  [1]{\csname bibitem#1\endcsname}%
\let\auto@bib@innerbib\@empty
\bibitem [{\citenamefont {Baltz}\ \emph {et~al.}(2018)\citenamefont {Baltz},
  \citenamefont {Manchon}, \citenamefont {Tsoi}, \citenamefont {Moriyama},
  \citenamefont {Ono},\ and\ \citenamefont
  {Tserkovnyak}}]{RevModPhys.90.015005}%
  \BibitemOpen
  \bibfield  {author} {\bibinfo {author} {\bibfnamefont {V.}~\bibnamefont
  {Baltz}}, \bibinfo {author} {\bibfnamefont {A.}~\bibnamefont {Manchon}},
  \bibinfo {author} {\bibfnamefont {M.}~\bibnamefont {Tsoi}}, \bibinfo {author}
  {\bibfnamefont {T.}~\bibnamefont {Moriyama}}, \bibinfo {author}
  {\bibfnamefont {T.}~\bibnamefont {Ono}},\ and\ \bibinfo {author}
  {\bibfnamefont {Y.}~\bibnamefont {Tserkovnyak}},\ }\bibfield  {title}
  {\bibinfo {title} {Antiferromagnetic spintronics},\ }\href
  {https://doi.org/10.1103/RevModPhys.90.015005} {\bibfield  {journal}
  {\bibinfo  {journal} {Rev. Mod. Phys.}\ }\textbf {\bibinfo {volume} {90}},\
  \bibinfo {pages} {015005} (\bibinfo {year} {2018})}\BibitemShut {NoStop}%
\bibitem [{\citenamefont {Nova}\ \emph {et~al.}(2017)\citenamefont {Nova},
  \citenamefont {Cartella}, \citenamefont {Cantaluppi}, \citenamefont
  {F{\"o}rst}, \citenamefont {Bossini}, \citenamefont {Mikhaylovskiy},
  \citenamefont {Kimel}, \citenamefont {Merlin},\ and\ \citenamefont
  {Cavalleri}}]{Nova_NP_2017}%
  \BibitemOpen
  \bibfield  {author} {\bibinfo {author} {\bibfnamefont {T.~F.}\ \bibnamefont
  {Nova}}, \bibinfo {author} {\bibfnamefont {A.}~\bibnamefont {Cartella}},
  \bibinfo {author} {\bibfnamefont {A.}~\bibnamefont {Cantaluppi}}, \bibinfo
  {author} {\bibfnamefont {M.}~\bibnamefont {F{\"o}rst}}, \bibinfo {author}
  {\bibfnamefont {D.}~\bibnamefont {Bossini}}, \bibinfo {author} {\bibfnamefont
  {R.~V.}\ \bibnamefont {Mikhaylovskiy}}, \bibinfo {author} {\bibfnamefont
  {A.~V.}\ \bibnamefont {Kimel}}, \bibinfo {author} {\bibfnamefont
  {R.}~\bibnamefont {Merlin}},\ and\ \bibinfo {author} {\bibfnamefont
  {A.}~\bibnamefont {Cavalleri}},\ }\bibfield  {title} {\bibinfo {title} {An
  effective magnetic field from optically driven phonons},\ }\href
  {https://doi.org/10.1038/nphys3925} {\bibfield  {journal} {\bibinfo
  {journal} {Nature Physics}\ }\textbf {\bibinfo {volume} {13}},\ \bibinfo
  {pages} {132} (\bibinfo {year} {2017})}\BibitemShut {NoStop}%
\bibitem [{\citenamefont {Seki}\ \emph {et~al.}(2015)\citenamefont {Seki},
  \citenamefont {Ideue}, \citenamefont {Kubota}, \citenamefont {Kozuka},
  \citenamefont {Takagi}, \citenamefont {Nakamura}, \citenamefont {Kaneko},
  \citenamefont {Kawasaki},\ and\ \citenamefont {Tokura}}]{Seki_PRL_2015}%
  \BibitemOpen
  \bibfield  {author} {\bibinfo {author} {\bibfnamefont {S.}~\bibnamefont
  {Seki}}, \bibinfo {author} {\bibfnamefont {T.}~\bibnamefont {Ideue}},
  \bibinfo {author} {\bibfnamefont {M.}~\bibnamefont {Kubota}}, \bibinfo
  {author} {\bibfnamefont {Y.}~\bibnamefont {Kozuka}}, \bibinfo {author}
  {\bibfnamefont {R.}~\bibnamefont {Takagi}}, \bibinfo {author} {\bibfnamefont
  {M.}~\bibnamefont {Nakamura}}, \bibinfo {author} {\bibfnamefont
  {Y.}~\bibnamefont {Kaneko}}, \bibinfo {author} {\bibfnamefont
  {M.}~\bibnamefont {Kawasaki}},\ and\ \bibinfo {author} {\bibfnamefont
  {Y.}~\bibnamefont {Tokura}},\ }\bibfield  {title} {\bibinfo {title} {Thermal
  generation of spin current in an antiferromagnet},\ }\href
  {https://link.aps.org/doi/10.1103/PhysRevLett.115.266601} {\bibfield
  {journal} {\bibinfo  {journal} {Phys. Rev. Lett.}\ }\textbf {\bibinfo
  {volume} {115}} (\bibinfo {year} {2015})}\BibitemShut {NoStop}%
\bibitem [{\citenamefont {Kravchuk}\ \emph {et~al.}(2018)\citenamefont
  {Kravchuk}, \citenamefont {Sheka}, \citenamefont {K\'akay}, \citenamefont
  {Volkov}, \citenamefont {R\"o\ss{}ler}, \citenamefont {van~den Brink},
  \citenamefont {Makarov},\ and\ \citenamefont {Gaididei}}]{Kravchuk_PRL2018}%
  \BibitemOpen
  \bibfield  {author} {\bibinfo {author} {\bibfnamefont {V.~P.}\ \bibnamefont
  {Kravchuk}}, \bibinfo {author} {\bibfnamefont {D.~D.}\ \bibnamefont {Sheka}},
  \bibinfo {author} {\bibfnamefont {A.}~\bibnamefont {K\'akay}}, \bibinfo
  {author} {\bibfnamefont {O.~M.}\ \bibnamefont {Volkov}}, \bibinfo {author}
  {\bibfnamefont {U.~K.}\ \bibnamefont {R\"o\ss{}ler}}, \bibinfo {author}
  {\bibfnamefont {J.}~\bibnamefont {van~den Brink}}, \bibinfo {author}
  {\bibfnamefont {D.}~\bibnamefont {Makarov}},\ and\ \bibinfo {author}
  {\bibfnamefont {Y.}~\bibnamefont {Gaididei}},\ }\bibfield  {title} {\bibinfo
  {title} {Multiplet of skyrmion states on a curvilinear defect: Reconfigurable
  skyrmion lattices},\ }\href {https://doi.org/10.1103/PhysRevLett.120.067201}
  {\bibfield  {journal} {\bibinfo  {journal} {Phys. Rev. Lett.}\ }\textbf
  {\bibinfo {volume} {120}},\ \bibinfo {pages} {067201} (\bibinfo {year}
  {2018})}\BibitemShut {NoStop}%
\bibitem [{\citenamefont {Narayan}\ \emph {et~al.}(2019)\citenamefont
  {Narayan}, \citenamefont {Cano}, \citenamefont {Balatsky},\ and\
  \citenamefont {Spaldin}}]{Narayan_NM_2019}%
  \BibitemOpen
  \bibfield  {author} {\bibinfo {author} {\bibfnamefont {A.}~\bibnamefont
  {Narayan}}, \bibinfo {author} {\bibfnamefont {A.}~\bibnamefont {Cano}},
  \bibinfo {author} {\bibfnamefont {A.~V.}\ \bibnamefont {Balatsky}},\ and\
  \bibinfo {author} {\bibfnamefont {N.~A.}\ \bibnamefont {Spaldin}},\
  }\bibfield  {title} {\bibinfo {title} {Multiferroic quantum criticality},\
  }\href {https://doi.org/10.1038/s41563-018-0255-6} {\bibfield  {journal}
  {\bibinfo  {journal} {Nature Materials}\ }\textbf {\bibinfo {volume} {18}},\
  \bibinfo {pages} {223} (\bibinfo {year} {2019})}\BibitemShut {NoStop}%
\bibitem [{\citenamefont {Chen}\ \emph {et~al.}(2019)\citenamefont {Chen},
  \citenamefont {Zhao}, \citenamefont {Artyukhin},\ and\ \citenamefont
  {Bellaiche}}]{Bellaiche_2020}%
  \BibitemOpen
  \bibfield  {author} {\bibinfo {author} {\bibfnamefont {P.}~\bibnamefont
  {Chen}}, \bibinfo {author} {\bibfnamefont {H.}~\bibnamefont {Zhao}}, \bibinfo
  {author} {\bibfnamefont {S.}~\bibnamefont {Artyukhin}},\ and\ \bibinfo
  {author} {\bibfnamefont {L.}~\bibnamefont {Bellaiche}},\ }\href@noop {}
  {\bibinfo {title} {{LINVARIANT}}},\ \bibinfo {howpublished}
  {\url{https://github.com/PaulChern/LINVARIANT}} (\bibinfo {year}
  {2019})\BibitemShut {NoStop}%
\bibitem [{\citenamefont {Ma}\ \emph {et~al.}(2008)\citenamefont {Ma},
  \citenamefont {Woo},\ and\ \citenamefont {Dudarev}}]{MPW_PRB2008}%
  \BibitemOpen
  \bibfield  {author} {\bibinfo {author} {\bibfnamefont {P.-W.}\ \bibnamefont
  {Ma}}, \bibinfo {author} {\bibfnamefont {C.~H.}\ \bibnamefont {Woo}},\ and\
  \bibinfo {author} {\bibfnamefont {S.~L.}\ \bibnamefont {Dudarev}},\
  }\bibfield  {title} {\bibinfo {title} {Large-scale simulation of the
  spin-lattice dynamics in ferromagnetic iron},\ }\href
  {https://doi.org/10.1103/PhysRevB.78.024434} {\bibfield  {journal} {\bibinfo
  {journal} {Phys. Rev. B}\ }\textbf {\bibinfo {volume} {78}},\ \bibinfo
  {pages} {024434} (\bibinfo {year} {2008})}\BibitemShut {NoStop}%
\bibitem [{\citenamefont {Ma}\ \emph {et~al.}(2016)\citenamefont {Ma},
  \citenamefont {Dudarev},\ and\ \citenamefont {Woo}}]{MPW_CPC2016}%
  \BibitemOpen
  \bibfield  {author} {\bibinfo {author} {\bibfnamefont {P.-W.}\ \bibnamefont
  {Ma}}, \bibinfo {author} {\bibfnamefont {S.~L.}\ \bibnamefont {Dudarev}},\
  and\ \bibinfo {author} {\bibfnamefont {C.~H.}\ \bibnamefont {Woo}},\
  }\bibfield  {title} {\bibinfo {title} {\ce{SPILADY}: A parallel \ce{CPU} and
  \ce{GPU} code for spin–lattice magnetic molecular dynamics simulations},\
  }\href {https://doi.org/10.1016/j.cpc.2016.05.017} {\bibfield  {journal}
  {\bibinfo  {journal} {Comput. Phys. Commun.}\ }\textbf {\bibinfo {volume}
  {207}},\ \bibinfo {pages} {350} (\bibinfo {year} {2016})}\BibitemShut
  {NoStop}%
\bibitem [{\citenamefont {Perera}\ \emph {et~al.}(2017)\citenamefont {Perera},
  \citenamefont {Nicholson}, \citenamefont {Eisenbach}, \citenamefont
  {Stocks},\ and\ \citenamefont {Landau}}]{PD_PRB2017}%
  \BibitemOpen
  \bibfield  {author} {\bibinfo {author} {\bibfnamefont {D.}~\bibnamefont
  {Perera}}, \bibinfo {author} {\bibfnamefont {D.~M.}\ \bibnamefont
  {Nicholson}}, \bibinfo {author} {\bibfnamefont {M.}~\bibnamefont
  {Eisenbach}}, \bibinfo {author} {\bibfnamefont {G.~M.}\ \bibnamefont
  {Stocks}},\ and\ \bibinfo {author} {\bibfnamefont {D.~P.}\ \bibnamefont
  {Landau}},\ }\bibfield  {title} {\bibinfo {title} {Collective dynamics in
  atomistic models with coupled translational and spin degrees of freedom},\
  }\href {https://doi.org/10.1103/PhysRevB.95.014431} {\bibfield  {journal}
  {\bibinfo  {journal} {Phys. Rev. B}\ }\textbf {\bibinfo {volume} {95}},\
  \bibinfo {pages} {014431} (\bibinfo {year} {2017})}\BibitemShut {NoStop}%
\bibitem [{\citenamefont {Tranchida}\ \emph {et~al.}(2018)\citenamefont
  {Tranchida}, \citenamefont {Plimpton}, \citenamefont {Thibaudeau},\ and\
  \citenamefont {Thompson}}]{tranchida2018massively}%
  \BibitemOpen
  \bibfield  {author} {\bibinfo {author} {\bibfnamefont {J.}~\bibnamefont
  {Tranchida}}, \bibinfo {author} {\bibfnamefont {S.~J.}\ \bibnamefont
  {Plimpton}}, \bibinfo {author} {\bibfnamefont {P.}~\bibnamefont
  {Thibaudeau}},\ and\ \bibinfo {author} {\bibfnamefont {A.~P.}\ \bibnamefont
  {Thompson}},\ }\bibfield  {title} {\bibinfo {title} {Massively parallel
  symplectic algorithm for coupled magnetic spin dynamics and molecular
  dynamics},\ }\href
  {https://doi.org/https://doi.org/10.1016/j.jcp.2018.06.042} {\bibfield
  {journal} {\bibinfo  {journal} {Journal of Computational Physics}\ }\textbf
  {\bibinfo {volume} {372}},\ \bibinfo {pages} {406} (\bibinfo {year}
  {2018})}\BibitemShut {NoStop}%
\bibitem [{\citenamefont {Nikolov}\ \emph {et~al.}(2021)\citenamefont
  {Nikolov}, \citenamefont {Wood}, \citenamefont {Cangi}, \citenamefont
  {Maillet}, \citenamefont {Marinica}, \citenamefont {Thompson}, \citenamefont
  {Desjarlais},\ and\ \citenamefont {Tranchida}}]{Nikolov2021}%
  \BibitemOpen
  \bibfield  {author} {\bibinfo {author} {\bibfnamefont {S.}~\bibnamefont
  {Nikolov}}, \bibinfo {author} {\bibfnamefont {M.~A.}\ \bibnamefont {Wood}},
  \bibinfo {author} {\bibfnamefont {A.}~\bibnamefont {Cangi}}, \bibinfo
  {author} {\bibfnamefont {J.-B.}\ \bibnamefont {Maillet}}, \bibinfo {author}
  {\bibfnamefont {M.-C.}\ \bibnamefont {Marinica}}, \bibinfo {author}
  {\bibfnamefont {A.~P.}\ \bibnamefont {Thompson}}, \bibinfo {author}
  {\bibfnamefont {M.~P.}\ \bibnamefont {Desjarlais}},\ and\ \bibinfo {author}
  {\bibfnamefont {J.}~\bibnamefont {Tranchida}},\ }\bibfield  {title} {\bibinfo
  {title} {Data-driven magneto-elastic predictions with scalable classical
  spin-lattice dynamics},\ }\href {https://doi.org/10.1038/s41524-021-00617-2}
  {\bibfield  {journal} {\bibinfo  {journal} {npj Computational Materials}\
  }\textbf {\bibinfo {volume} {7}},\ \bibinfo {pages} {153} (\bibinfo {year}
  {2021})}\BibitemShut {NoStop}%
\bibitem [{\citenamefont {Fechner}\ \emph {et~al.}(2018)\citenamefont
  {Fechner}, \citenamefont {Sukhov}, \citenamefont {Chotorlishvili},
  \citenamefont {Kenel}, \citenamefont {Berakdar},\ and\ \citenamefont
  {Spaldin}}]{MFechner_PRB_2018}%
  \BibitemOpen
  \bibfield  {author} {\bibinfo {author} {\bibfnamefont {M.}~\bibnamefont
  {Fechner}}, \bibinfo {author} {\bibfnamefont {A.}~\bibnamefont {Sukhov}},
  \bibinfo {author} {\bibfnamefont {L.}~\bibnamefont {Chotorlishvili}},
  \bibinfo {author} {\bibfnamefont {C.}~\bibnamefont {Kenel}}, \bibinfo
  {author} {\bibfnamefont {J.}~\bibnamefont {Berakdar}},\ and\ \bibinfo
  {author} {\bibfnamefont {N.~A.}\ \bibnamefont {Spaldin}},\ }\bibfield
  {title} {\bibinfo {title} {Magnetophononics: Ultrafast spin control through
  the lattice},\ }\href {https://doi.org/10.1103/PhysRevMaterials.2.064401}
  {\bibfield  {journal} {\bibinfo  {journal} {Phys. Rev. Mater.}\ }\textbf
  {\bibinfo {volume} {2}},\ \bibinfo {pages} {064401} (\bibinfo {year}
  {2018})}\BibitemShut {NoStop}%
\bibitem [{\citenamefont {Zhang}\ \emph
  {et~al.}(2018{\natexlab{a}})\citenamefont {Zhang}, \citenamefont {Han},
  \citenamefont {Wang}, \citenamefont {Car},\ and\ \citenamefont
  {E}}]{ZLF_PRL2018}%
  \BibitemOpen
  \bibfield  {author} {\bibinfo {author} {\bibfnamefont {L.}~\bibnamefont
  {Zhang}}, \bibinfo {author} {\bibfnamefont {J.}~\bibnamefont {Han}}, \bibinfo
  {author} {\bibfnamefont {H.}~\bibnamefont {Wang}}, \bibinfo {author}
  {\bibfnamefont {R.}~\bibnamefont {Car}},\ and\ \bibinfo {author}
  {\bibfnamefont {W.}~\bibnamefont {E}},\ }\bibfield  {title} {\bibinfo {title}
  {Deep potential molecular dynamics: A scalable model with the accuracy of
  quantum mechanics},\ }\href {https://doi.org/10.1103/PhysRevLett.120.143001}
  {\bibfield  {journal} {\bibinfo  {journal} {Phys. Rev. Lett.}\ }\textbf
  {\bibinfo {volume} {120}},\ \bibinfo {pages} {143001} (\bibinfo {year}
  {2018}{\natexlab{a}})}\BibitemShut {NoStop}%
\bibitem [{\citenamefont {Li}\ \emph {et~al.}(2023)\citenamefont {Li},
  \citenamefont {Tang}, \citenamefont {Gong}, \citenamefont {Zou},
  \citenamefont {Duan},\ and\ \citenamefont {Xu}}]{li2022deeplearning}%
  \BibitemOpen
  \bibfield  {author} {\bibinfo {author} {\bibfnamefont {H.}~\bibnamefont
  {Li}}, \bibinfo {author} {\bibfnamefont {Z.}~\bibnamefont {Tang}}, \bibinfo
  {author} {\bibfnamefont {X.}~\bibnamefont {Gong}}, \bibinfo {author}
  {\bibfnamefont {N.}~\bibnamefont {Zou}}, \bibinfo {author} {\bibfnamefont
  {W.}~\bibnamefont {Duan}},\ and\ \bibinfo {author} {\bibfnamefont
  {Y.}~\bibnamefont {Xu}},\ }\bibfield  {title} {\bibinfo {title}
  {Deep-learning electronic-structure calculation of magnetic
  superstructures},\ }\href {https://doi.org/10.1038/s43588-023-00424-3}
  {\bibfield  {journal} {\bibinfo  {journal} {Nature Computational Science}\
  }\textbf {\bibinfo {volume} {3}},\ \bibinfo {pages} {321} (\bibinfo {year}
  {2023})}\BibitemShut {NoStop}%
\bibitem [{\citenamefont {Novikov}\ \emph {et~al.}(2022)\citenamefont
  {Novikov}, \citenamefont {Grabowski}, \citenamefont {K{\"o}rmann},\ and\
  \citenamefont {Shapeev}}]{novikov2022magnetic}%
  \BibitemOpen
  \bibfield  {author} {\bibinfo {author} {\bibfnamefont {I.}~\bibnamefont
  {Novikov}}, \bibinfo {author} {\bibfnamefont {B.}~\bibnamefont {Grabowski}},
  \bibinfo {author} {\bibfnamefont {F.}~\bibnamefont {K{\"o}rmann}},\ and\
  \bibinfo {author} {\bibfnamefont {A.}~\bibnamefont {Shapeev}},\ }\bibfield
  {title} {\bibinfo {title} {Magnetic moment tensor potentials for collinear
  spin-polarized materials reproduce different magnetic states of bcc
  \ce{Fe}},\ }\href {https://doi.org/10.1038/s41524-022-00696-9} {\bibfield
  {journal} {\bibinfo  {journal} {npj Computational Materials}\ }\textbf
  {\bibinfo {volume} {8}},\ \bibinfo {pages} {13} (\bibinfo {year}
  {2022})}\BibitemShut {NoStop}%
\bibitem [{\citenamefont {Zhang}\ \emph
  {et~al.}(2018{\natexlab{b}})\citenamefont {Zhang}, \citenamefont {Han},
  \citenamefont {Wang}, \citenamefont {Car},\ and\ \citenamefont
  {E}}]{zhang2018deep}%
  \BibitemOpen
  \bibfield  {author} {\bibinfo {author} {\bibfnamefont {L.}~\bibnamefont
  {Zhang}}, \bibinfo {author} {\bibfnamefont {J.}~\bibnamefont {Han}}, \bibinfo
  {author} {\bibfnamefont {H.}~\bibnamefont {Wang}}, \bibinfo {author}
  {\bibfnamefont {R.}~\bibnamefont {Car}},\ and\ \bibinfo {author}
  {\bibfnamefont {W.}~\bibnamefont {E}},\ }\bibfield  {title} {\bibinfo {title}
  {{Deep Potential Molecular Dynamics: A Scalable Model with the Accuracy of
  Quantum Mechanics}},\ }\href {https://doi.org/10.1103/PhysRevLett.120.143001}
  {\bibfield  {journal} {\bibinfo  {journal} {Phys. Rev. Lett.}\ }\textbf
  {\bibinfo {volume} {120}},\ \bibinfo {pages} {143001} (\bibinfo {year}
  {2018}{\natexlab{b}})}\BibitemShut {NoStop}%
\bibitem [{\citenamefont {Zhang}\ \emph
  {et~al.}(2018{\natexlab{c}})\citenamefont {Zhang}, \citenamefont {Han},
  \citenamefont {Wang}, \citenamefont {Saidi}, \citenamefont {Car},\ and\
  \citenamefont {E}}]{zhang2018end}%
  \BibitemOpen
  \bibfield  {author} {\bibinfo {author} {\bibfnamefont {L.}~\bibnamefont
  {Zhang}}, \bibinfo {author} {\bibfnamefont {J.}~\bibnamefont {Han}}, \bibinfo
  {author} {\bibfnamefont {H.}~\bibnamefont {Wang}}, \bibinfo {author}
  {\bibfnamefont {W.}~\bibnamefont {Saidi}}, \bibinfo {author} {\bibfnamefont
  {R.}~\bibnamefont {Car}},\ and\ \bibinfo {author} {\bibfnamefont
  {W.}~\bibnamefont {E}},\ }\bibfield  {title} {\bibinfo {title} {{End-to-end
  Symmetry Preserving Inter-atomic Potential Energy Model for Finite and
  Extended Systems}},\ }in\ \href
  {https://proceedings.neurips.cc/paper_files/paper/2018/file/e2ad76f2326fbc6b56a45a56c59fafdb-Paper.pdf}
  {\emph {\bibinfo {booktitle} {Advances in Neural Information Processing
  Systems}}},\ Vol.~\bibinfo {volume} {31},\ \bibinfo {editor} {edited by\
  \bibinfo {editor} {\bibfnamefont {S.}~\bibnamefont {Bengio}}, \bibinfo
  {editor} {\bibfnamefont {H.}~\bibnamefont {Wallach}}, \bibinfo {editor}
  {\bibfnamefont {H.}~\bibnamefont {Larochelle}}, \bibinfo {editor}
  {\bibfnamefont {K.}~\bibnamefont {Grauman}}, \bibinfo {editor} {\bibfnamefont
  {N.}~\bibnamefont {Cesa-Bianchi}},\ and\ \bibinfo {editor} {\bibfnamefont
  {R.}~\bibnamefont {Garnett}}}\ (\bibinfo  {publisher} {Curran Associates,
  Inc.},\ \bibinfo {year} {2018})\BibitemShut {NoStop}%
\bibitem [{\citenamefont {Wang}\ \emph {et~al.}(2018)\citenamefont {Wang},
  \citenamefont {Zhang}, \citenamefont {Han},\ and\ \citenamefont
  {E}}]{wang2018deepmd}%
  \BibitemOpen
  \bibfield  {author} {\bibinfo {author} {\bibfnamefont {H.}~\bibnamefont
  {Wang}}, \bibinfo {author} {\bibfnamefont {L.}~\bibnamefont {Zhang}},
  \bibinfo {author} {\bibfnamefont {J.}~\bibnamefont {Han}},\ and\ \bibinfo
  {author} {\bibfnamefont {W.}~\bibnamefont {E}},\ }\bibfield  {title}
  {\bibinfo {title} {{DeePMD-kit: A deep learning package for many-body
  potential energy representation and molecular dynamics}},\ }\href
  {https://doi.org/https://doi.org/10.1016/j.cpc.2018.03.016} {\bibfield
  {journal} {\bibinfo  {journal} {Computer Physics Communications}\ }\textbf
  {\bibinfo {volume} {228}},\ \bibinfo {pages} {178} (\bibinfo {year}
  {2018})}\BibitemShut {NoStop}%
\bibitem [{\citenamefont {Goodfellow}\ \emph {et~al.}(2016)\citenamefont
  {Goodfellow}, \citenamefont {Bengio},\ and\ \citenamefont
  {Courville}}]{goodfellow2016deep}%
  \BibitemOpen
  \bibfield  {author} {\bibinfo {author} {\bibfnamefont {I.~J.}\ \bibnamefont
  {Goodfellow}}, \bibinfo {author} {\bibfnamefont {Y.}~\bibnamefont {Bengio}},\
  and\ \bibinfo {author} {\bibfnamefont {A.}~\bibnamefont {Courville}},\
  }\href@noop {} {\emph {\bibinfo {title} {{Deep Learning}}}}\ (\bibinfo
  {publisher} {MIT press},\ \bibinfo {address} {Cambridge, MA, USA},\ \bibinfo
  {year} {2016})\BibitemShut {NoStop}%
\bibitem [{\citenamefont {He}\ \emph {et~al.}(2016)\citenamefont {He},
  \citenamefont {Zhang}, \citenamefont {Ren},\ and\ \citenamefont
  {Sun}}]{he2016deep}%
  \BibitemOpen
  \bibfield  {author} {\bibinfo {author} {\bibfnamefont {K.}~\bibnamefont
  {He}}, \bibinfo {author} {\bibfnamefont {X.}~\bibnamefont {Zhang}}, \bibinfo
  {author} {\bibfnamefont {S.}~\bibnamefont {Ren}},\ and\ \bibinfo {author}
  {\bibfnamefont {J.}~\bibnamefont {Sun}},\ }\bibfield  {title} {\bibinfo
  {title} {{Deep Residual Learning for Image Recognition}},\ }in\ \href
  {https://openaccess.thecvf.com/content_cvpr_2016/papers/He_Deep_Residual_Learning_CVPR_2016_paper.pdf}
  {\emph {\bibinfo {booktitle} {Proceedings of the IEEE Conference on Computer
  Vision and Pattern Recognition (CVPR)}}}\ (\bibinfo {year}
  {2016})\BibitemShut {NoStop}%
\bibitem [{\citenamefont {Kingma}\ and\ \citenamefont
  {Ba}(2017)}]{kingma2017adam}%
  \BibitemOpen
  \bibfield  {author} {\bibinfo {author} {\bibfnamefont {D.~P.}\ \bibnamefont
  {Kingma}}\ and\ \bibinfo {author} {\bibfnamefont {J.}~\bibnamefont {Ba}},\
  }\href {https://arxiv.org/pdf/1412.6980.pdf} {\bibinfo {title} {{Adam: A
  Method for Stochastic Optimization}}} (\bibinfo {year} {2017}),\ \Eprint
  {https://arxiv.org/abs/1412.6980} {arXiv:1412.6980 [cs.LG]} \BibitemShut
  {NoStop}%
\bibitem [{\citenamefont {Cai}\ \emph {et~al.}(2023)\citenamefont {Cai},
  \citenamefont {Wang}, \citenamefont {Xu}, \citenamefont {Wei},\ and\
  \citenamefont {Xu}}]{cai2023firstprinciples}%
  \BibitemOpen
  \bibfield  {author} {\bibinfo {author} {\bibfnamefont {Z.}~\bibnamefont
  {Cai}}, \bibinfo {author} {\bibfnamefont {K.}~\bibnamefont {Wang}}, \bibinfo
  {author} {\bibfnamefont {Y.}~\bibnamefont {Xu}}, \bibinfo {author}
  {\bibfnamefont {S.-H.}\ \bibnamefont {Wei}},\ and\ \bibinfo {author}
  {\bibfnamefont {B.}~\bibnamefont {Xu}},\ }\href
  {https://arxiv.org/pdf/2208.04551.pdf} {\bibinfo {title} {First-principles
  study of non-collinear spin fluctuations using self-adaptive spin-constrained
  method}} (\bibinfo {year} {2023}),\ \Eprint
  {https://arxiv.org/abs/2208.04551} {arXiv:2208.04551 [cond-mat.mtrl-sci]}
  \BibitemShut {NoStop}%
\bibitem [{\citenamefont {Feynman}(1939)}]{feynman1939forces}%
  \BibitemOpen
  \bibfield  {author} {\bibinfo {author} {\bibfnamefont {R.~P.}\ \bibnamefont
  {Feynman}},\ }\bibfield  {title} {\bibinfo {title} {{Forces in Molecules}},\
  }\href {https://doi.org/10.1103/PhysRev.56.340} {\bibfield  {journal}
  {\bibinfo  {journal} {Phys. Rev.}\ }\textbf {\bibinfo {volume} {56}},\
  \bibinfo {pages} {340} (\bibinfo {year} {1939})}\BibitemShut {NoStop}%
\bibitem [{\citenamefont {Zhang}\ \emph {et~al.}(2019)\citenamefont {Zhang},
  \citenamefont {Lin}, \citenamefont {Wang}, \citenamefont {Car},\ and\
  \citenamefont {E}}]{zhang2019active}%
  \BibitemOpen
  \bibfield  {author} {\bibinfo {author} {\bibfnamefont {L.}~\bibnamefont
  {Zhang}}, \bibinfo {author} {\bibfnamefont {D.-Y.}\ \bibnamefont {Lin}},
  \bibinfo {author} {\bibfnamefont {H.}~\bibnamefont {Wang}}, \bibinfo {author}
  {\bibfnamefont {R.}~\bibnamefont {Car}},\ and\ \bibinfo {author}
  {\bibfnamefont {W.}~\bibnamefont {E}},\ }\bibfield  {title} {\bibinfo {title}
  {{Active learning of uniformly accurate interatomic potentials for materials
  simulation}},\ }\href {https://doi.org/10.1103/PhysRevMaterials.3.023804}
  {\bibfield  {journal} {\bibinfo  {journal} {Phys. Rev. Mater.}\ }\textbf
  {\bibinfo {volume} {3}},\ \bibinfo {pages} {023804} (\bibinfo {year}
  {2019})}\BibitemShut {NoStop}%
\bibitem [{\citenamefont {Zhang}\ \emph {et~al.}(2020)\citenamefont {Zhang},
  \citenamefont {Wang}, \citenamefont {Chen}, \citenamefont {Zeng},
  \citenamefont {Zhang}, \citenamefont {Wang},\ and\ \citenamefont
  {E}}]{zhang2020dpgen}%
  \BibitemOpen
  \bibfield  {author} {\bibinfo {author} {\bibfnamefont {Y.}~\bibnamefont
  {Zhang}}, \bibinfo {author} {\bibfnamefont {H.}~\bibnamefont {Wang}},
  \bibinfo {author} {\bibfnamefont {W.}~\bibnamefont {Chen}}, \bibinfo {author}
  {\bibfnamefont {J.}~\bibnamefont {Zeng}}, \bibinfo {author} {\bibfnamefont
  {L.}~\bibnamefont {Zhang}}, \bibinfo {author} {\bibfnamefont
  {H.}~\bibnamefont {Wang}},\ and\ \bibinfo {author} {\bibfnamefont
  {W.}~\bibnamefont {E}},\ }\bibfield  {title} {\bibinfo {title} {{DP-GEN: A
  concurrent learning platform for the generation of reliable deep learning
  based potential energy models}},\ }\href
  {https://doi.org/https://doi.org/10.1016/j.cpc.2020.107206} {\bibfield
  {journal} {\bibinfo  {journal} {Computer Physics Communications}\ }\textbf
  {\bibinfo {volume} {253}},\ \bibinfo {pages} {107206} (\bibinfo {year}
  {2020})}\BibitemShut {NoStop}%
\bibitem [{\citenamefont {Milano}\ and\ \citenamefont
  {Grimsditch}(2010)}]{milano2010magnetic}%
  \BibitemOpen
  \bibfield  {author} {\bibinfo {author} {\bibfnamefont {J.}~\bibnamefont
  {Milano}}\ and\ \bibinfo {author} {\bibfnamefont {M.}~\bibnamefont
  {Grimsditch}},\ }\bibfield  {title} {\bibinfo {title} {{Magnetic field
  effects on the NiO magnon spectra}},\ }\href
  {https://doi.org/10.1103/PhysRevB.81.094415} {\bibfield  {journal} {\bibinfo
  {journal} {Phys. Rev. B}\ }\textbf {\bibinfo {volume} {81}},\ \bibinfo
  {pages} {094415} (\bibinfo {year} {2010})}\BibitemShut {NoStop}%
\bibitem [{\citenamefont {Massey}\ \emph {et~al.}(1990)\citenamefont {Massey},
  \citenamefont {Chen}, \citenamefont {Allen},\ and\ \citenamefont
  {Merlin}}]{massey1990pressure}%
  \BibitemOpen
  \bibfield  {author} {\bibinfo {author} {\bibfnamefont {M.~J.}\ \bibnamefont
  {Massey}}, \bibinfo {author} {\bibfnamefont {N.~H.}\ \bibnamefont {Chen}},
  \bibinfo {author} {\bibfnamefont {J.~W.}\ \bibnamefont {Allen}},\ and\
  \bibinfo {author} {\bibfnamefont {R.}~\bibnamefont {Merlin}},\ }\bibfield
  {title} {\bibinfo {title} {{Pressure dependence of two-magnon Raman
  scattering in NiO}},\ }\href {https://doi.org/10.1103/PhysRevB.42.8776}
  {\bibfield  {journal} {\bibinfo  {journal} {Phys. Rev. B}\ }\textbf {\bibinfo
  {volume} {42}},\ \bibinfo {pages} {8776} (\bibinfo {year}
  {1990})}\BibitemShut {NoStop}%
\bibitem [{\citenamefont {Aytan}\ \emph {et~al.}(2017)\citenamefont {Aytan},
  \citenamefont {Debnath}, \citenamefont {Kargar}, \citenamefont {Barlas},
  \citenamefont {Lacerda}, \citenamefont {Li}, \citenamefont {Lake},
  \citenamefont {Shi},\ and\ \citenamefont {Balandin}}]{aytan2017spin}%
  \BibitemOpen
  \bibfield  {author} {\bibinfo {author} {\bibfnamefont {E.}~\bibnamefont
  {Aytan}}, \bibinfo {author} {\bibfnamefont {B.}~\bibnamefont {Debnath}},
  \bibinfo {author} {\bibfnamefont {F.}~\bibnamefont {Kargar}}, \bibinfo
  {author} {\bibfnamefont {Y.}~\bibnamefont {Barlas}}, \bibinfo {author}
  {\bibfnamefont {M.~M.}\ \bibnamefont {Lacerda}}, \bibinfo {author}
  {\bibfnamefont {J.~X.}\ \bibnamefont {Li}}, \bibinfo {author} {\bibfnamefont
  {R.~K.}\ \bibnamefont {Lake}}, \bibinfo {author} {\bibfnamefont
  {J.}~\bibnamefont {Shi}},\ and\ \bibinfo {author} {\bibfnamefont {A.~A.}\
  \bibnamefont {Balandin}},\ }\bibfield  {title} {\bibinfo {title}
  {{Spin-phonon coupling in antiferromagnetic nickel oxide}},\ }\href
  {https://doi.org/10.1063/1.5009598} {\bibfield  {journal} {\bibinfo
  {journal} {Applied Physics Letters}\ }\textbf {\bibinfo {volume} {111}},\
  \bibinfo {pages} {252402} (\bibinfo {year} {2017})}\BibitemShut {NoStop}%
\bibitem [{\citenamefont {Fischer}\ \emph {et~al.}(1980)\citenamefont
  {Fischer}, \citenamefont {Polomska}, \citenamefont {Sosnowska},\ and\
  \citenamefont {Szymanski}}]{fischer1980temperature}%
  \BibitemOpen
  \bibfield  {author} {\bibinfo {author} {\bibfnamefont {P.}~\bibnamefont
  {Fischer}}, \bibinfo {author} {\bibfnamefont {M.}~\bibnamefont {Polomska}},
  \bibinfo {author} {\bibfnamefont {I.}~\bibnamefont {Sosnowska}},\ and\
  \bibinfo {author} {\bibfnamefont {M.}~\bibnamefont {Szymanski}},\ }\bibfield
  {title} {\bibinfo {title} {{Temperature dependence of the crystal and
  magnetic structures of BiFeO$_3$}},\ }\href
  {https://doi.org/10.1088/0022-3719/13/10/012} {\bibfield  {journal} {\bibinfo
   {journal} {Journal of Physics C: Solid State Physics}\ }\textbf {\bibinfo
  {volume} {13}},\ \bibinfo {pages} {1931} (\bibinfo {year}
  {1980})}\BibitemShut {NoStop}%
\bibitem [{\citenamefont {Ederer}\ and\ \citenamefont
  {Spaldin}(2005)}]{ederer2005weak}%
  \BibitemOpen
  \bibfield  {author} {\bibinfo {author} {\bibfnamefont {C.}~\bibnamefont
  {Ederer}}\ and\ \bibinfo {author} {\bibfnamefont {N.~A.}\ \bibnamefont
  {Spaldin}},\ }\bibfield  {title} {\bibinfo {title} {{Weak ferromagnetism and
  magnetoelectric coupling in bismuth ferrite}},\ }\href
  {https://doi.org/10.1103/PhysRevB.71.060401} {\bibfield  {journal} {\bibinfo
  {journal} {Phys. Rev. B}\ }\textbf {\bibinfo {volume} {71}},\ \bibinfo
  {pages} {060401} (\bibinfo {year} {2005})}\BibitemShut {NoStop}%
\bibitem [{\citenamefont {Albrecht}\ \emph {et~al.}(2010)\citenamefont
  {Albrecht}, \citenamefont {Lisenkov}, \citenamefont {Ren}, \citenamefont
  {Rahmedov}, \citenamefont {Kornev},\ and\ \citenamefont
  {Bellaiche}}]{albrecht2010ferromagnetism}%
  \BibitemOpen
  \bibfield  {author} {\bibinfo {author} {\bibfnamefont {D.}~\bibnamefont
  {Albrecht}}, \bibinfo {author} {\bibfnamefont {S.}~\bibnamefont {Lisenkov}},
  \bibinfo {author} {\bibfnamefont {W.}~\bibnamefont {Ren}}, \bibinfo {author}
  {\bibfnamefont {D.}~\bibnamefont {Rahmedov}}, \bibinfo {author}
  {\bibfnamefont {I.~A.}\ \bibnamefont {Kornev}},\ and\ \bibinfo {author}
  {\bibfnamefont {L.}~\bibnamefont {Bellaiche}},\ }\bibfield  {title} {\bibinfo
  {title} {{Ferromagnetism in multiferroic ${\text{BiFeO}}_{3}$ films: A
  first-principles-based study}},\ }\href
  {https://doi.org/10.1103/PhysRevB.81.140401} {\bibfield  {journal} {\bibinfo
  {journal} {Phys. Rev. B}\ }\textbf {\bibinfo {volume} {81}},\ \bibinfo
  {pages} {140401} (\bibinfo {year} {2010})}\BibitemShut {NoStop}%
\bibitem [{\citenamefont {Xu}\ \emph {et~al.}(2021)\citenamefont {Xu},
  \citenamefont {Meyer}, \citenamefont {Verstraete}, \citenamefont
  {Bellaiche},\ and\ \citenamefont {Dup\'e}}]{xu2021first}%
  \BibitemOpen
  \bibfield  {author} {\bibinfo {author} {\bibfnamefont {B.}~\bibnamefont
  {Xu}}, \bibinfo {author} {\bibfnamefont {S.}~\bibnamefont {Meyer}}, \bibinfo
  {author} {\bibfnamefont {M.~J.}\ \bibnamefont {Verstraete}}, \bibinfo
  {author} {\bibfnamefont {L.}~\bibnamefont {Bellaiche}},\ and\ \bibinfo
  {author} {\bibfnamefont {B.}~\bibnamefont {Dup\'e}},\ }\bibfield  {title}
  {\bibinfo {title} {{First-principles study of spin spirals in the
  multiferroic ${\mathrm{BiFeO}}_{3}$}},\ }\href
  {https://doi.org/10.1103/PhysRevB.103.214423} {\bibfield  {journal} {\bibinfo
   {journal} {Phys. Rev. B}\ }\textbf {\bibinfo {volume} {103}},\ \bibinfo
  {pages} {214423} (\bibinfo {year} {2021})}\BibitemShut {NoStop}%
\bibitem [{\citenamefont {Vignale}\ and\ \citenamefont
  {Kohn}(1996)}]{vignale1996current}%
  \BibitemOpen
  \bibfield  {author} {\bibinfo {author} {\bibfnamefont {G.}~\bibnamefont
  {Vignale}}\ and\ \bibinfo {author} {\bibfnamefont {W.}~\bibnamefont {Kohn}},\
  }\bibfield  {title} {\bibinfo {title} {{Current-Dependent
  Exchange-Correlation Potential for Dynamical Linear Response Theory}},\
  }\href {https://doi.org/10.1103/PhysRevLett.77.2037} {\bibfield  {journal}
  {\bibinfo  {journal} {Phys. Rev. Lett.}\ }\textbf {\bibinfo {volume} {77}},\
  \bibinfo {pages} {2037} (\bibinfo {year} {1996})}\BibitemShut {NoStop}%
\bibitem [{\citenamefont {Qian}\ and\ \citenamefont
  {Vignale}(2002)}]{qian2002dynamical}%
  \BibitemOpen
  \bibfield  {author} {\bibinfo {author} {\bibfnamefont {Z.}~\bibnamefont
  {Qian}}\ and\ \bibinfo {author} {\bibfnamefont {G.}~\bibnamefont {Vignale}},\
  }\bibfield  {title} {\bibinfo {title} {{Dynamical exchange-correlation
  potentials for an electron liquid}},\ }\href
  {https://doi.org/10.1103/PhysRevB.65.235121} {\bibfield  {journal} {\bibinfo
  {journal} {Phys. Rev. B}\ }\textbf {\bibinfo {volume} {65}},\ \bibinfo
  {pages} {235121} (\bibinfo {year} {2002})}\BibitemShut {NoStop}%
\bibitem [{\citenamefont {Gilbert}(2004)}]{gilbert2004phenomenological}%
  \BibitemOpen
  \bibfield  {author} {\bibinfo {author} {\bibfnamefont {T.}~\bibnamefont
  {Gilbert}},\ }\bibfield  {title} {\bibinfo {title} {{A phenomenological
  theory of damping in ferromagnetic materials}},\ }\href
  {https://doi.org/10.1109/TMAG.2004.836740} {\bibfield  {journal} {\bibinfo
  {journal} {IEEE Transactions on Magnetics}\ }\textbf {\bibinfo {volume}
  {40}},\ \bibinfo {pages} {3443} (\bibinfo {year} {2004})}\BibitemShut
  {NoStop}%
\bibitem [{\citenamefont {Kumar}\ \emph {et~al.}(2012)\citenamefont {Kumar},
  \citenamefont {Fennie},\ and\ \citenamefont {Rabe}}]{kumar2012spin}%
  \BibitemOpen
  \bibfield  {author} {\bibinfo {author} {\bibfnamefont {A.}~\bibnamefont
  {Kumar}}, \bibinfo {author} {\bibfnamefont {C.~J.}\ \bibnamefont {Fennie}},\
  and\ \bibinfo {author} {\bibfnamefont {K.~M.}\ \bibnamefont {Rabe}},\
  }\bibfield  {title} {\bibinfo {title} {{Spin-lattice coupling and phonon
  dispersion of CdCr${}_{2}$O${}_{4}$ from first principles}},\ }\href
  {https://doi.org/10.1103/PhysRevB.86.184429} {\bibfield  {journal} {\bibinfo
  {journal} {Phys. Rev. B}\ }\textbf {\bibinfo {volume} {86}},\ \bibinfo
  {pages} {184429} (\bibinfo {year} {2012})}\BibitemShut {NoStop}%
\bibitem [{\citenamefont {Wu}\ \emph {et~al.}(2018)\citenamefont {Wu},
  \citenamefont {Liu},\ and\ \citenamefont {Luo}}]{wu2018magnon}%
  \BibitemOpen
  \bibfield  {author} {\bibinfo {author} {\bibfnamefont {X.}~\bibnamefont
  {Wu}}, \bibinfo {author} {\bibfnamefont {Z.}~\bibnamefont {Liu}},\ and\
  \bibinfo {author} {\bibfnamefont {T.}~\bibnamefont {Luo}},\ }\bibfield
  {title} {\bibinfo {title} {{Magnon and phonon dispersion, lifetime, and
  thermal conductivity of iron from spin-lattice dynamics simulations}},\
  }\href {https://doi.org/10.1063/1.5020611} {\bibfield  {journal} {\bibinfo
  {journal} {Journal of Applied Physics}\ }\textbf {\bibinfo {volume} {123}},\
  \bibinfo {pages} {085109} (\bibinfo {year} {2018})}\BibitemShut {NoStop}%
\bibitem [{\citenamefont {Hellsvik}\ \emph {et~al.}(2019)\citenamefont
  {Hellsvik}, \citenamefont {Thonig}, \citenamefont {Modin}, \citenamefont
  {Iu\ifmmode~\mbox{\c{s}}\else \c{s}\fi{}an}, \citenamefont {Bergman},
  \citenamefont {Eriksson}, \citenamefont {Bergqvist},\ and\ \citenamefont
  {Delin}}]{hellsvik2019general}%
  \BibitemOpen
  \bibfield  {author} {\bibinfo {author} {\bibfnamefont {J.}~\bibnamefont
  {Hellsvik}}, \bibinfo {author} {\bibfnamefont {D.}~\bibnamefont {Thonig}},
  \bibinfo {author} {\bibfnamefont {K.}~\bibnamefont {Modin}}, \bibinfo
  {author} {\bibfnamefont {D.}~\bibnamefont {Iu\ifmmode~\mbox{\c{s}}\else
  \c{s}\fi{}an}}, \bibinfo {author} {\bibfnamefont {A.}~\bibnamefont
  {Bergman}}, \bibinfo {author} {\bibfnamefont {O.}~\bibnamefont {Eriksson}},
  \bibinfo {author} {\bibfnamefont {L.}~\bibnamefont {Bergqvist}},\ and\
  \bibinfo {author} {\bibfnamefont {A.}~\bibnamefont {Delin}},\ }\bibfield
  {title} {\bibinfo {title} {{General method for atomistic spin-lattice
  dynamics with first-principles accuracy}},\ }\href
  {https://doi.org/10.1103/PhysRevB.99.104302} {\bibfield  {journal} {\bibinfo
  {journal} {Phys. Rev. B}\ }\textbf {\bibinfo {volume} {99}},\ \bibinfo
  {pages} {104302} (\bibinfo {year} {2019})}\BibitemShut {NoStop}%
\bibitem [{\citenamefont {Merkle}\ and\ \citenamefont
  {Smith}(1987)}]{merkle1987atomic}%
  \BibitemOpen
  \bibfield  {author} {\bibinfo {author} {\bibfnamefont {K.~L.}\ \bibnamefont
  {Merkle}}\ and\ \bibinfo {author} {\bibfnamefont {D.~J.}\ \bibnamefont
  {Smith}},\ }\bibfield  {title} {\bibinfo {title} {{Atomic Structure of
  Symmetric Tilt Grain Boundaries in NiO}},\ }\href
  {https://doi.org/10.1103/PhysRevLett.59.2887} {\bibfield  {journal} {\bibinfo
   {journal} {Phys. Rev. Lett.}\ }\textbf {\bibinfo {volume} {59}},\ \bibinfo
  {pages} {2887} (\bibinfo {year} {1987})}\BibitemShut {NoStop}%
\bibitem [{\citenamefont {Li}\ \emph {et~al.}(2006)\citenamefont {Li},
  \citenamefont {Chen}, \citenamefont {Qihe},\ and\ \citenamefont
  {Li}}]{li2006magnetic}%
  \BibitemOpen
  \bibfield  {author} {\bibinfo {author} {\bibfnamefont {L.}~\bibnamefont
  {Li}}, \bibinfo {author} {\bibfnamefont {L.}~\bibnamefont {Chen}}, \bibinfo
  {author} {\bibfnamefont {R.}~\bibnamefont {Qihe}},\ and\ \bibinfo {author}
  {\bibfnamefont {G.}~\bibnamefont {Li}},\ }\bibfield  {title} {\bibinfo
  {title} {{Magnetic crossover of NiO nanocrystals at room temperature}},\
  }\href {https://doi.org/10.1063/1.2357562} {\bibfield  {journal} {\bibinfo
  {journal} {Applied Physics Letters}\ }\textbf {\bibinfo {volume} {89}},\
  \bibinfo {pages} {134102} (\bibinfo {year} {2006})}\BibitemShut {NoStop}%
\bibitem [{\citenamefont {Henkelman}\ \emph {et~al.}(2000)\citenamefont
  {Henkelman}, \citenamefont {Uberuaga},\ and\ \citenamefont
  {Jónsson}}]{henkelman2000climbing}%
  \BibitemOpen
  \bibfield  {author} {\bibinfo {author} {\bibfnamefont {G.}~\bibnamefont
  {Henkelman}}, \bibinfo {author} {\bibfnamefont {B.~P.}\ \bibnamefont
  {Uberuaga}},\ and\ \bibinfo {author} {\bibfnamefont {H.}~\bibnamefont
  {Jónsson}},\ }\bibfield  {title} {\bibinfo {title} {{A climbing image nudged
  elastic band method for finding saddle points and minimum energy paths}},\
  }\href {https://doi.org/10.1063/1.1329672} {\bibfield  {journal} {\bibinfo
  {journal} {The Journal of Chemical Physics}\ }\textbf {\bibinfo {volume}
  {113}},\ \bibinfo {pages} {9901} (\bibinfo {year} {2000})}\BibitemShut
  {NoStop}%
\bibitem [{\citenamefont {Munro}\ \emph {et~al.}(2018)\citenamefont {Munro},
  \citenamefont {Akamatsu}, \citenamefont {Padmanabhan}, \citenamefont {Liu},
  \citenamefont {Shi}, \citenamefont {Chen}, \citenamefont {VanLeeuwen},
  \citenamefont {Dabo},\ and\ \citenamefont {Gopalan}}]{munro2018discovering}%
  \BibitemOpen
  \bibfield  {author} {\bibinfo {author} {\bibfnamefont {J.~M.}\ \bibnamefont
  {Munro}}, \bibinfo {author} {\bibfnamefont {H.}~\bibnamefont {Akamatsu}},
  \bibinfo {author} {\bibfnamefont {H.}~\bibnamefont {Padmanabhan}}, \bibinfo
  {author} {\bibfnamefont {V.~S.}\ \bibnamefont {Liu}}, \bibinfo {author}
  {\bibfnamefont {Y.}~\bibnamefont {Shi}}, \bibinfo {author} {\bibfnamefont
  {L.-Q.}\ \bibnamefont {Chen}}, \bibinfo {author} {\bibfnamefont {B.~K.}\
  \bibnamefont {VanLeeuwen}}, \bibinfo {author} {\bibfnamefont
  {I.}~\bibnamefont {Dabo}},\ and\ \bibinfo {author} {\bibfnamefont
  {V.}~\bibnamefont {Gopalan}},\ }\bibfield  {title} {\bibinfo {title}
  {{Discovering minimum energy pathways via distortion symmetry groups}},\
  }\href {https://doi.org/10.1103/PhysRevB.98.085107} {\bibfield  {journal}
  {\bibinfo  {journal} {Phys. Rev. B}\ }\textbf {\bibinfo {volume} {98}},\
  \bibinfo {pages} {085107} (\bibinfo {year} {2018})}\BibitemShut {NoStop}%
\bibitem [{\citenamefont {Zavaliche}\ \emph {et~al.}(2005)\citenamefont
  {Zavaliche}, \citenamefont {Shafer}, \citenamefont {Ramesh}, \citenamefont
  {Cruz}, \citenamefont {Das}, \citenamefont {Kim},\ and\ \citenamefont
  {Eom}}]{zavaliche2005polarization}%
  \BibitemOpen
  \bibfield  {author} {\bibinfo {author} {\bibfnamefont {F.}~\bibnamefont
  {Zavaliche}}, \bibinfo {author} {\bibfnamefont {P.}~\bibnamefont {Shafer}},
  \bibinfo {author} {\bibfnamefont {R.}~\bibnamefont {Ramesh}}, \bibinfo
  {author} {\bibfnamefont {M.~P.}\ \bibnamefont {Cruz}}, \bibinfo {author}
  {\bibfnamefont {R.~R.}\ \bibnamefont {Das}}, \bibinfo {author} {\bibfnamefont
  {D.~M.}\ \bibnamefont {Kim}},\ and\ \bibinfo {author} {\bibfnamefont {C.~B.}\
  \bibnamefont {Eom}},\ }\bibfield  {title} {\bibinfo {title} {{Polarization
  switching in epitaxial BiFeO3 films}},\ }\href
  {https://doi.org/10.1063/1.2149180} {\bibfield  {journal} {\bibinfo
  {journal} {Applied Physics Letters}\ }\textbf {\bibinfo {volume} {87}},\
  \bibinfo {pages} {252902} (\bibinfo {year} {2005})}\BibitemShut {NoStop}%
\bibitem [{\citenamefont {Heron}\ \emph {et~al.}(2014)\citenamefont {Heron},
  \citenamefont {Bosse}, \citenamefont {He}, \citenamefont {Gao}, \citenamefont
  {Trassin}, \citenamefont {Ye}, \citenamefont {Clarkson}, \citenamefont
  {Wang}, \citenamefont {Liu}, \citenamefont {Salahuddin} \emph
  {et~al.}}]{heron2014deterministic}%
  \BibitemOpen
  \bibfield  {author} {\bibinfo {author} {\bibfnamefont {J.}~\bibnamefont
  {Heron}}, \bibinfo {author} {\bibfnamefont {J.}~\bibnamefont {Bosse}},
  \bibinfo {author} {\bibfnamefont {Q.}~\bibnamefont {He}}, \bibinfo {author}
  {\bibfnamefont {Y.}~\bibnamefont {Gao}}, \bibinfo {author} {\bibfnamefont
  {M.}~\bibnamefont {Trassin}}, \bibinfo {author} {\bibfnamefont
  {L.}~\bibnamefont {Ye}}, \bibinfo {author} {\bibfnamefont {J.}~\bibnamefont
  {Clarkson}}, \bibinfo {author} {\bibfnamefont {C.}~\bibnamefont {Wang}},
  \bibinfo {author} {\bibfnamefont {J.}~\bibnamefont {Liu}}, \bibinfo {author}
  {\bibfnamefont {S.}~\bibnamefont {Salahuddin}}, \emph {et~al.},\ }\bibfield
  {title} {\bibinfo {title} {{Deterministic switching of ferromagnetism at room
  temperature using an electric field}},\ }\href
  {https://doi.org/10.1038/nature14004} {\bibfield  {journal} {\bibinfo
  {journal} {Nature}\ }\textbf {\bibinfo {volume} {516}},\ \bibinfo {pages}
  {370} (\bibinfo {year} {2014})}\BibitemShut {NoStop}%
\end{thebibliography}%

\end{document}